\documentclass[sn-aps]{sn-jnl}

\usepackage{graphicx}%
\usepackage{multirow}%
\usepackage{amsmath,amssymb,amsfonts}%
\usepackage{amsthm}%
\usepackage{mathrsfs}%
\usepackage[title]{appendix}%
\usepackage{xcolor}%
\usepackage{textcomp}%
\usepackage{manyfoot}%
\usepackage{booktabs}%
\usepackage{algorithm}%
\usepackage{algorithmicx}%
\usepackage{algpseudocode}%
\usepackage{listings}%
\usepackage{lmodern}
\usepackage{geometry}
\usepackage{indentfirst}

\theoremstyle{thmstyleone}%
\theoremstyle{thmstyletwo}%

\theoremstyle{thmstylethree}%

\begin{document}

\title[Article Title]{Quantum-preserved transport of excitations in Rydberg-dressed atom arrays} 


\author[1]{\fnm{Panpan} \sur{Li} }

\author*[1,2,4,5]{\fnm{Jing} \sur{Qian} }\email{jqian@phy.ecnu.edu.cn}

\author[3,4,5,6]{ \fnm{Weiping} \sur{Zhang}}

\affil[1]{\orgdiv{School of Physics and Electronic Science},\orgname{ East China Normal University},  \orgaddress{\city{Shanghai}, \postcode{200062},  \country{China}}}

\affil[2]{\orgdiv{Chongqing Institute of East China Normal University}, \orgaddress{ \city{Chongqing}, \postcode{401120}, \country{China}}}

\affil[3]{\orgdiv{School of Physics and Astronomy, and Tsung-Dao Lee Institute}, \orgname{Shanghai Jiao Tong University}, \orgaddress{ \city{Shanghai}, \postcode{200240}, \country{China}}}

\affil[4]{\orgdiv{Shanghai Research Center for Quantum Science}, \orgname{Chinese Academy of Sciences, Shanghai Municipal Government}, \orgaddress{ \city{Shanghai}, \postcode{201315}, \country{China}}}

\affil[5]{\orgdiv{Shanghai Branch, Hefei National Laboratory}, \orgaddress{\city{Shanghai}, \postcode{201315}, \country{China}}}

\affil[6]{\orgdiv{Collaborative Innovation Center of Extreme Optics}, \orgname{Shanxi University}, \orgaddress{ \city{Taiyuan}, \postcode{030006}, \country{China}}}

\abstract{To transport high-quality quantum state between two distant qubits through
one-dimensional spin chains, the perfect state transfer (PST) method serves as the first choice, due to its natively perfect transfer fidelity that is independent of the system dimension. However, the PST requires a precise modulation of the local pulse parameters as well as an accurate timing of dynamic evolution, and is thus very sensitive to variations in practice. Here, we propose a protocol for achieving quantum-preserved transport of excitations using an array of Rydberg-dressed atoms, enabled by optimal control of minimally global parameters. By treating the weak coupling of two marginal array atoms as a perturbation, an effective spin-exchange model with highly tunable interactions between the external weak and the inner strong driving atoms can be established, which allows for coherent excitation transfer even with large atomic position fluctuation. We furthermore show that the existence of long-time excitation propagation unattainable for systems under antiblockade facilitation conditions.
Our results highlight an easily-implemented scheme for studying the dynamics of spin systems using Rydberg atoms and may guide the avenue to the engineering of complex many-body dynamics.}


\maketitle


\newpage

\section{Introduction}\label{sec1}
For long-distance quantum communication a high-quality quantum state transport has been considered as the most fundamental task \textcolor{black}{\cite{Nature.404.247,Nat.Photon.8.356,Nature.Photon.9.641}}. E.g. it is crucial to transfer information with high fidelity between two distant qubits in large-scale fault-tolerant quantum computers \textcolor{black}{\cite{Nature.598.281,Nature.549.172,PRX.Quantum.3.010329}}.
Common ways of quantum state transport depend on a precise engineering of the special couplings between the nearest-neighbor (NN) sites, known as perfect state transfer (PST) \textcolor{black}{\cite{Phys.Rev.Lett.92.187902}}, yet which adds experimental difficulty due to the multiple local pulse controls \textcolor{black}{\cite{PhysRevA.93.012343}}, and meanwhile is very sensitive to practical variations such as the atomic position fluctuation \textcolor{black}{\cite{PhysRevA.73.032306}}. Applying well-designed external fields \textcolor{black}{\cite{PhysRevA.71.032309,PhysRevA.77.012303,PhysRevA.82.022318,PhysRevA.88.062309}} or coupling strengths \textcolor{black}{\cite{NewJ.Phys.9.155,PhysRevA.97.052333,PhysRevResearch.2.033475,PhysRevA.108.032401}} may solve this problem, for instance, in shortcut to adiabaticity (STA) protocols,
one could use an auxiliarily counterdiabatic modulation to achieve a fast and robust state transport by simply suppressing unwanted transitions \textcolor{black}{\cite{PhysRevLett.111.100502,PhysRevA.96.021801,PhysRevA.105.052422,Phys.Rev.A.97.012333}}. However, this auxiliary driving strategy usually needs a complex form and remains difficult to realize \textcolor{black}{\cite{PhysRevLett.123.090602,Science.Bulletin}}. Since the PST cannot be achieved by fully uniform couplings, the potential viability of perfect transport through a non-uniform coupled atomic array using minimally global controls is becoming increasingly attractive.

So far the spin excitation encoded by an array of laser-dressed Rydberg atoms treats as a good candidate for such task which provides a promising platform for studies of quantum simulation and quantum information processing \textcolor{black}{\cite{PhysRevLett.120.180502,Nature.534.66, Science.373.1359,RevModPhys.82.2313}}. In a Rydberg quantum simulator, pseudo-spin can be mapped into different Rydberg manifolds allowing for implementing various spin-$1/2$ models \textcolor{black}{\cite{PhysRevLett.120.063601,PhysRevLett.130.243001,PhysRevLett.114.113002}}, where a direct spin-exchange interaction can be induced featuring both high tunability and long-range properties. Earlier efforts for realizing such spin-exchange interaction could utilize two
highly-excited Rydberg states mediated by resonance dipole-dipole interactions \textcolor{black}{\cite{ PhysRevA.104.063301,nat.phys.10.914}}, one ground state and one Rydberg state by a second-order process in terms of laser-dressed couplings \textcolor{black}{\cite{PhysRevA.97.043415,PhysRevLett.115.093002}}, as well as the ground-state manifolds through second-order \textcolor{black}{\cite{PhysRevA.105.032417}} and four-order processes \textcolor{black}{\cite{PhysRevLett.114.173002,PhysRevLett.114.243002}}. For instance, Yang {\it et.al.} \textcolor{black}{\cite{PhysRevLett.123.063001}} derived a simple and effective model for engineering the excitation transport dynamics where a pure synthetic spin-exchange arises from diagonal van der Waals (vdWs) interaction avoiding the use of resonant dipolar interaction or Floquet engineering with multiple Rydberg levels \textcolor{black}{\cite{PhysRevA.108.053318,PRXQuantum.3.020303}}. Whilst, the local PST condition still impedes this model to be used for a robust quantum state transport based on the state-of-the-art experimental technologies.

Inspired by this work, we put forward an efficient protocol for transporting excitation in an one-dimensional (1D) Rydberg-atom array with only control of the two marginal couplings. When the boundary couplings are rather weak as compared to the uniform and large intermediate atomic couplings, known as the weak-coupling (WC) condition \textcolor{black}{\cite{PhysRevB.82.140403,PhysRevA.71.022301}}, the dynamic evolution of system can persist in the Zeno subspace formed by merely three eigenstates $\{|\Phi_1\rangle,|\Phi_m\rangle,|\Phi_N\rangle\}$ regardless of the system dimension, which is called quantum Zeno dynamics \textcolor{black}{\cite{PhysRevLett.89.080401}}. We show that, by precisely varying the external large detunings which provide a handle to change the relative energy of three eigenstates in the Zeno subspace, a quantum-preserved excitation transport exerted on two marginal array atoms can be realized. 
To investigate the robustness of transport dynamics we introduce the model with strong atomic position fluctuation arising from finite temperature, and finally, by making use of minimally optimized parameters we successfully achieve a robust Rydberg excitation transport with high fidelity benefiting from a strong insensitivity to both the atomic position uncertainty \textcolor{black}{\cite{PhysRevA.101.043421}} as well as the dephasing error \textcolor{black}{\cite{PhysRevA.99.043404}}. This work may serve as an important step to the exploration of complex spin-model dynamics in Rydberg quantum simulation \textcolor{black}{\cite{Nature.616.691}}.

\section{Theoretical strategy}\label{sec2}

\begin{figure}
\centering
\includegraphics[width=0.6\textwidth]{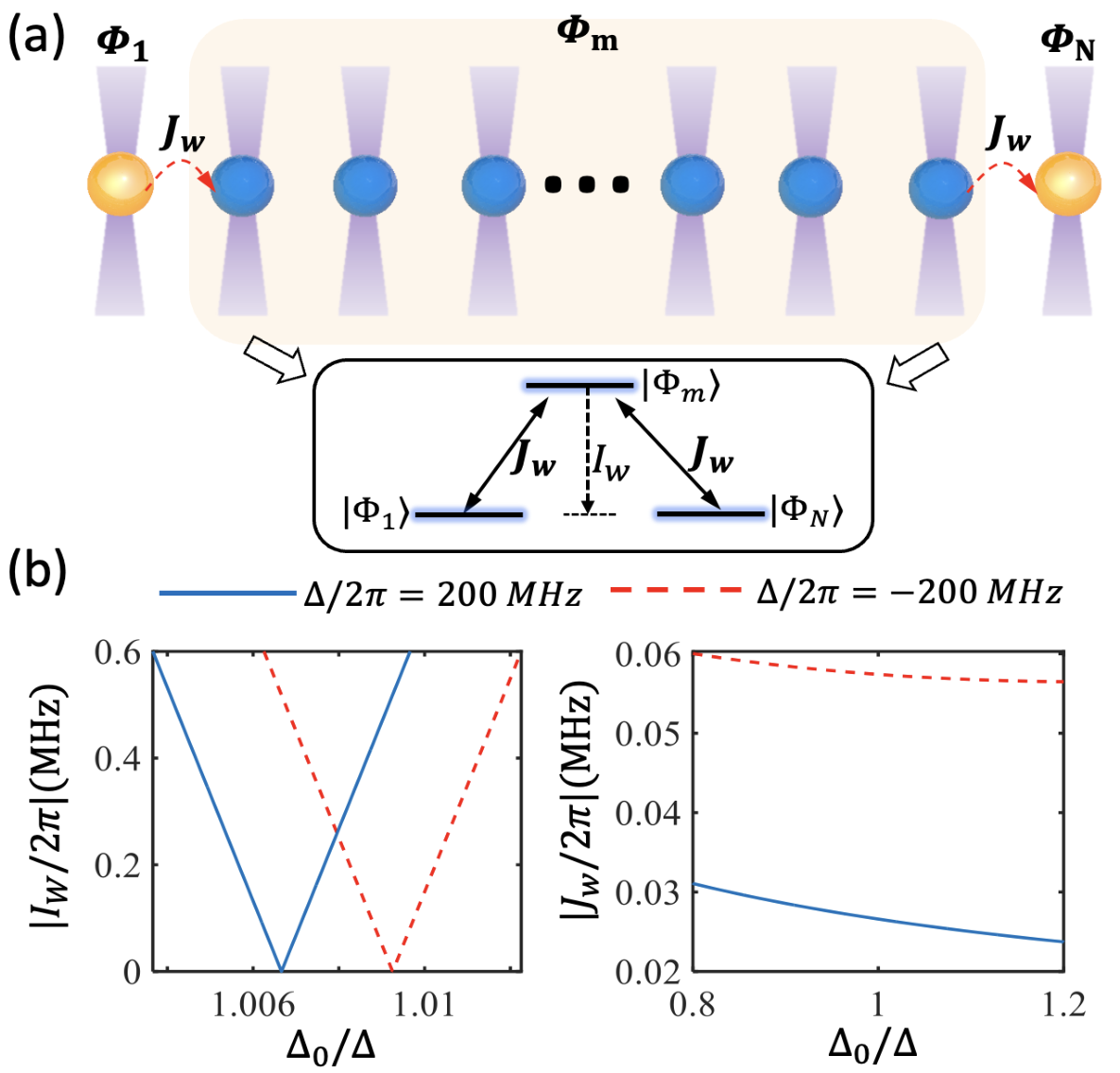}
\caption{Schematic of a quantum-preserved excitation transport based on a 1D Rydberg array using the WC condition. (a) Constructing an effective model by
a chain of off-resonant excited atoms within the $Z=\{|\Phi_1\rangle,|\Phi_m\rangle,|\Phi_N\rangle\}$ subspace manifesting as a three-level system (see box), where the coupling strength of two marginal atoms are weaker than that of the intermediate atoms, i.e. $\Omega_0\ll\Omega$. (b) The variation of $(I_W,J_W)$ as a function of the detuning ratio $\Delta_0/\Delta$ where $(\Omega_0,\Omega)/2\pi= (1,10)$ MHz and $N=4$. It is noted that the effective on-site potential $|I_W|$ can vanish at $\Delta_0/\Delta = C$ ($C$ is an uncertain number) for any sign of detuning $\Delta$.
\textcolor{black}{The specific values considered here are $C_6/2\pi =1416$ GHz$\cdot \mu$m$^6$, $r = 3.76$ $\mu$m (the NN distance), so that $V_{NN}/2\pi = 501 $ MHz, $V_{NNN}/2\pi = 7.83 $ MHz, proposed for $^{87}$Rb atoms using internal levels $|g_j\rangle= |5S_{1/2},F=1,m_F=0\rangle$ and $|r_j\rangle = |73S_{1/2}\rangle$ from the ARC source library }\textcolor{black}{\cite{RN31}.}
}
\label{fig1:nengji}
\end{figure}

\subsection{Description of the effective model}\label{subsec1}
 The system we consider, as shown in Fig.\ref{fig1:nengji}(a), is composed by an array of individually trapped cold atoms, dressed by far off-resonant laser fields which couple the ground state $|g_j\rangle$ to a Rydberg state $|r_j\rangle$ with local Rabi frequency $\Omega_j$ and local detuning $\Delta_j$, satisfying $\Delta_j\gg\Omega_j$. Such a scenario can be easily modeled by the original Hamiltonian \textcolor{black}{\cite{Nat.Phys.16.132}}
\begin{equation}
H=\sum_{j}\Omega_j \sigma_x^j+ \sum_j \Delta_j\sigma_{rr}^j
+\sum_{j<k} V_{jk} \sigma_{rr}^j\sigma_{rr}^k \label{ham}
\end{equation}
where $j,k$ are the array indices, $\sigma_x^j = |r_j\rangle\langle g_j|+|g_j\rangle\langle r_j|$, $\sigma_{rr}^j = |r_j\rangle\langle r_j|$ are respectively the transition and projection operators for the $j$th atom and $V_{jk} = C_6/r_{jk}^6$ is the vdWs interaction between atoms located at $\bold{r}_j$ and $\bold{r}_k$ with $r_{jk} = |\bold{r}_j-\bold{r}_k|$ the distance. The goal is to transfer Rydberg excitation between two marginal atoms of the array. To our knowledge, a high-quality quantum state transport plays a central role in the field of quantum communication \textcolor{black}{\cite{Contemp.Phys.48.13}}.
However, previous schemes (such as the PST) depend on a precise control of local parameters $(\Omega_j,\Delta_j)$ which is much sensitive to practical variations \textcolor{black}{\cite{PhysRevA.84.022311}}. Thus, to realize a high-fidelity and robust state transfer, we apply two marginal atoms at $\bold{r}_1$ and $\bold{r}_N$ with weak couplings as compared to that of the intermediate atoms, which is so-called the WC condition i.e. $\Omega_0\ll \Omega$ (note that $\Omega_0=\Omega_{1,N}$, $\Omega = \Omega_{2,...,N-1}$, similarly, $\Delta_0=\Delta_{1,N}$, $\Delta = \Delta_{2,...,N-1}$) \textcolor{black}{\cite{Phys.Rev.A.95.012317}}. That means an individual operation of every array atom by modulating all local parameters $\Omega_j$ and $\Delta_j$, is unnecessary. Only four global parameters $(\Omega_0,\Delta_0,\Omega,\Delta)$ need to be determined. 

Therefore, the effective Hamiltonian under the WC condition can be derived as (see Appendix \ref{secA1} for details)
\begin{eqnarray}
H_{WC,N}=  \left[\begin{array}{ccc}
     0&J_W&0 \\
     J_W&I_W&J_W   \\
     0&J_W&0 \label{wccn}
\end{array}\right]
\end{eqnarray}
based on the $Z=\{|\Phi_1\rangle,|\Phi_m\rangle,|\Phi_N\rangle\}$ subspace, same as Eq.(\ref{wcc}). Here, the exchange coupling strength $J_W$ and the effective Ising-type potential $I_W$ take explicit forms in (\ref{JIw}). Fig.\ref{fig1:nengji}(b) presents the explicit dependence of $I_W$ and $J_W$ by varying the marginal detuning $\Delta_0$ ($\Delta$ is fixed). Note that, \textcolor{black}{$\Delta$ is the detuning of two-photon transition and a negative $\Delta$ may cause a facilitated dynamics when $\Delta+V_{NN}=0$ (see Fig.\ref{fig4:error}e).}
For any sign of detuning we find $I_W$ always exists a zero point $|I_W| = 0$ at
$\Delta_0/\Delta=C$, beyond which the diagonal energy shift $|I_W|$ reveals a dramatic increase, manifesting one order of magnitude larger than the level of off-diagonal coupling $|J_W|$. Meanwhile, $|J_W|$ has a distinct trend featuring relatively steady with $\Delta_0$, and a negative detuning makes the coupling strength larger due to the partial compensation of the positive blockaded strengths $V_{NN},V_{NNN}$.
From the view of a reduced three-level model (see box in Fig.\ref{fig1:nengji}a), if $|I_W|\gg |J_W|$ the state transfer of $|\Phi_1\rangle\to|\Phi_N\rangle$ becomes a second-order process whose contribution scales as $\sim J_W^2/I_W$ providing a suppression of intermediate state $|\Phi_m\rangle$ in an adiabatic way \textcolor{black}{\cite{PhysRevA.84.022330}}. This typically requires a long evolution time $\sim \pi I_W/(2J_W^2)$
together with a low transfer fidelity due to the effect of dephasing error \textcolor{black}{\cite{PhysRevA.85.032111}}.
Only if $|I_W|\approx 0$ the state transfer can stay within the $Z$ subspace featuring a more efficient excitation transport.

\begin{figure}
\centering
\includegraphics[width=0.6\textwidth]{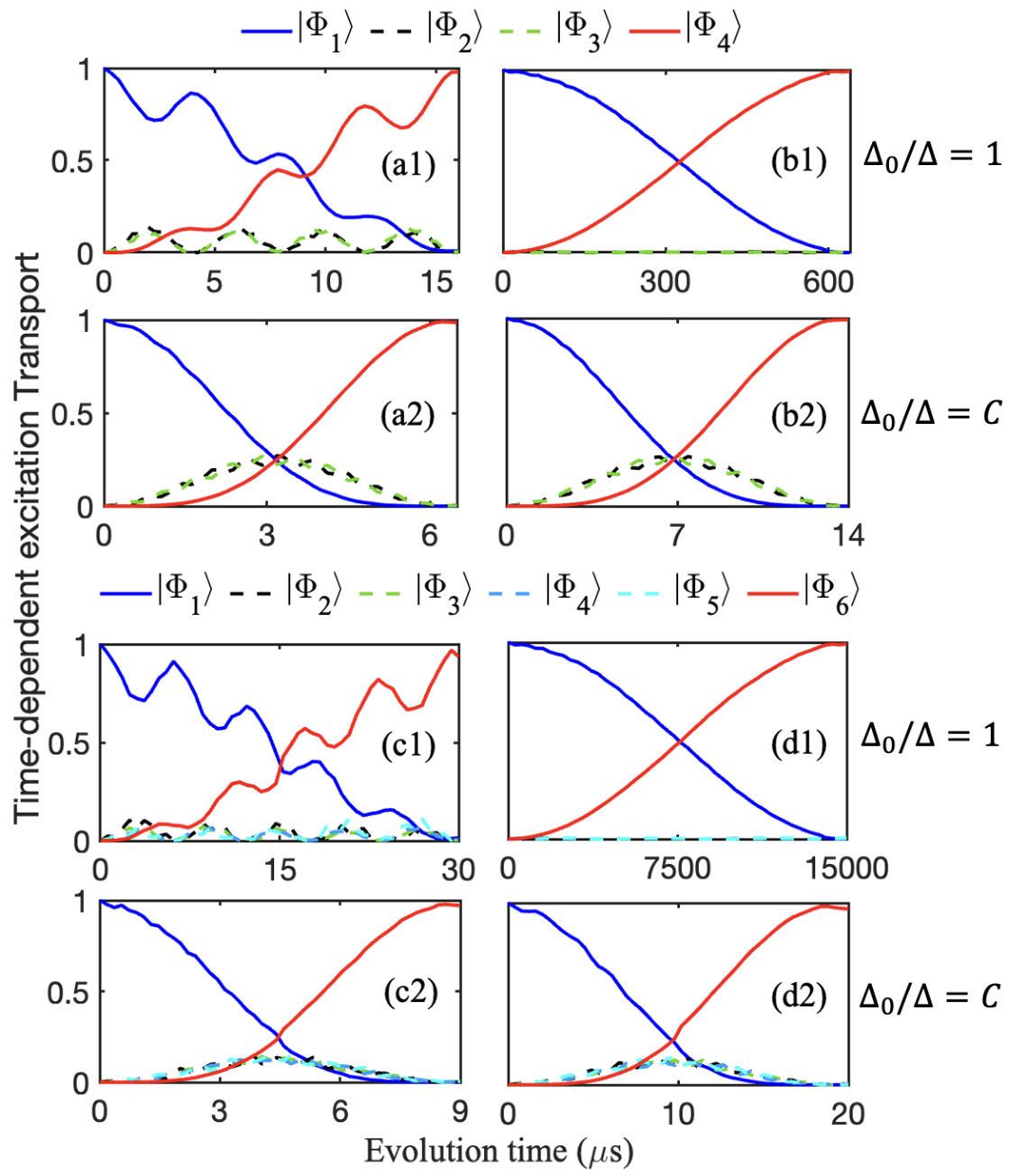}
\caption{Ideal time-dependent excitation transport without dephasing $\Gamma=0$ for $N=4$ (a1-2; b1-2) and $N=6$ (c1-2; d1-2), governed by the full Hamiltonian (\ref{ham}). Here, global parameters are taken as $\Delta/2\pi = -200$ MHz (left column) and 200 MHz (right column) and others are the same as in Fig.\ref{fig1:nengji}(b). $C=(1.0092,1.0066,1.0091,1.0066)$ for (a2;b2;c2;d2).}
\label{fig2:wci}
\end{figure}

\subsection{Numerical verification}\label{subsec2}
 To benchmark the excitation transport efficiency between two marginal atoms, we now know that, the choice of ratio $\Delta_0/\Delta$ (see Fig.\ref{fig1:nengji}b) is sensitively important which determines the relationship between $J_W$ and $I_W$. We verify this by performing an exact numerical integration of the full Hamiltonian (\ref{ham}), governed by the Lindblad master equation for the density matrix $\rho$ \textcolor{black}{\cite{ Phys.Rev.A.95.012708}}
\begin{equation}
   \dot{\rho} = -i[H,\rho] + \mathcal{L}[\rho] 
   \label{rrh}
\end{equation}
where the Lindblad superoperator $\mathcal{L}[\rho]$ describes the intrinsic dephasing error (denoted by rate $\Gamma$) caused by e.g. the laser phase noise \textcolor{black}{\cite{PhysRevA.97.053803}} as well as the spontaneous decay error (denoted by rate $\gamma$) from state $|r_j\rangle$, and thus 
\begin{equation}
\mathcal{L}[\rho] = \sum_{l=1,2}\sum_{j}[\mathcal L_{lj} \rho \mathcal L_{ lj}^\dagger -\frac{1}{2}(\mathcal L_{ lj}^\dagger \mathcal L_{lj} \rho+\rho \mathcal L_{lj}^\dagger \mathcal L_{lj})]     
\end{equation}
with \textcolor{black}{$\mathcal{L}_{1j} = \sqrt{\Gamma}(\sigma_{rr}^j - \sigma_{gg}^j) ,\mathcal{L}_{2j} =  \sqrt{\gamma}|g_j\rangle\langle r_j|$.} Accounting for that a typical Rydberg lifetime of $\sim 100$ $\mu$s \textcolor{black}{\cite{Phys.Rev.A.5.052504}}, which is much longer than the duration for excitation transport we ignore the decay error (i.e. $\gamma=0$) here. Finally, by solving the evolution of $\rho(t)$ with the original Hamiltonian $H$, the full (not effective) dynamics could be resolved, which conserves the total excitation number 1 within the single excitation subspace $\Pi_1$. The quality of state transfer is measured by the fidelity {$F(t) = \langle\Phi_N|\rho(t)|\Phi_N\rangle $} meaning the real-time local probability of state $|\Phi_N\rangle$ and $F$ presents the first maximum of fidelity after an one-time evolution at $t=T_g$ between two marginal atoms.

Figure \ref{fig2:wci} depicts the numerical results of transferring the initial excitation in $|\Phi_1(0)\rangle$ for $N=4$ and $6$ in the absence of dephasing error, $\Gamma=0$. In principle, we can raise the prefect fidelity to about $F_0\approx 1.0$ with arbitrary detuning as long as the WC condition $\Omega_0\ll\Omega$ is satisfied. Here, we choose $\Omega_0/\Omega=1/10$. However, if $\Delta_0/\Delta = 1$ as in Fig.\ref{fig2:wci}(a1) and (c1), the dynamical behavior exhibits strong oscillations for a negative detuning. \textcolor{black}{Because the on-site energy $|I_W|$ of $|\Phi_m\rangle$ and the off-diagonal coupling strength $|J_W|$ become comparable in this case leading to a near-resonance excitation transfer along with oscillations.} \textcolor{black}{E.g. for the (a1) case, $|I_W|/2\pi\approx0.1954$ MHz which is only 4 times larger than $|J_W|/2\pi\approx 0.0574$ MHz then the dynamics is affected by both competing terms. }While, if the detuning is positive, we find a near-perfect fidelity can be achieved yet at the cost of a much longer $T_g$ [see Fig.\ref{fig2:wci}(b1;d1)], especially for a larger $N$. For example, the time for the case of $N=4$ is $T_g\approx 640$ $\mu$s. But it will need more time $T_g\approx 15000$ $\mu$s to ensure the transfer for $N=6$. This distinct behavior agrees with the understanding of a second-order adiabatic process with an almost full suppression of the intermediate population due to $|I_W|\gg |J_W|$ \textcolor{black}{\cite{PhysRevA.85.032111}}. Up to now, we have shown two protocols for a high-fidelity transport in the WC picture; however, both of them are unsatisfactory: one (a1;c1) suffers from strong oscillations which is susceptible to the intrinsic dephasing and systematic errors and the other (b1;d1) needs a quite long time impacted by the decay error.

What's more, 
since at $\Delta_0/\Delta=C$ the system is effectively formed by quasi-degenerate three levels $\{\Phi_1\rangle,|\Phi_m\rangle,|\Phi_N\rangle\}$, it is instructive to look for more efficient excitation transports in this regime. The calculated full dynamics of single Rydberg excitation is displayed in (a2-b2) and (c2-d2) of Fig.\ref{fig2:wci} for $\Delta_0/\Delta=C$. We observe that a perfect excitation transfer between $|\Phi_1\rangle$ and $|\Phi_N\rangle$ which qualitatively agrees with the effective model with the expected evolution time, given by 
\begin{equation}
    T_g \approx \pi/(\sqrt{2}|J_W|) \label{tgg}
\end{equation}

Clearly, owing to the vanishing of $I_W$, the evolution time is solely dominated  by the coupling strength $|J_W|$. It reveals that a smaller $|J_W|$ leads to the increase of $T_g$. Such a trend is confirmed by comparing the results for $N=4$ and $6$. Accounting for $J_W \propto
 (N-2)^{-1/2}$, as $N$ increases (c2;d2), we find similar behavior as (a2) and (b2)
 except for a longer time. In fact there exists a trade-off between the coupling strength $J_W$ which affects the evolution time $T_g$ and the transfer fidelity $F$. A larger $|J_W|$ could shorten time alongside with a small excitation leakage. However, a larger $|J_W|$ may be at the cost of large Rabi frequencies which also breaks the off-resonance condition in the WC regime, allowing the whole transfer beyond the single excitation subspace.

\section{Optimal control strategy}\label{sec3}

\begin{table}
\caption{\label{table1:pulse}In Opt cases, the optimized parameters $(\Omega_0,\Omega,\Delta_0,\Delta)$ (in unit of $2\pi\times$MHz), the calculated values $|J_W|,|I_W|$ (in unit of $2\pi\times$MHz), the transfer fidelity $F$ in the presence of dephasing error ($F_0$ is the ideal value) and the evolution time $T_g$ (in unit of $\mu$s) based on a $N$-atom array. For a fair comparison, two non-optimal cases (labeled by Non-Opt and Mod) are given, correspondingly.
 }

\renewcommand{\arraystretch}{1.3}
\setlength{\tabcolsep}{0mm}{
\begin{tabular}{c@{\hspace{10pt}}c@{\hspace{10pt}}c@{\hspace{10pt}}c@{\hspace{10pt}}c@{\hspace{10pt}}c@{\hspace{10pt}}c@{\hspace{10pt}}c@{\hspace{10pt}}c@{\hspace{10pt}}c@{\hspace{10pt}}c}
\hline
Non-Opt Case &$N$ & $\Omega_0$ & $\Omega$ & $\Delta_0$ & $\Delta$  & $|J_W|$ & $|I_W|$  &$F_0$& $F$ & $T_g$ \\
\hline
I & 4 &1 & 10 & -200 &-200 & 0.0574 & 0.1954 &0.9839 &0.9188&15.9 \\
 & &1 & 10 &  200& 200 & 0.0266 & 1.8645  &0.9980& 0.3940 & 640.0 \\
\hline
 II & 4 &5 & 10 & -200 &-200 & 0.2870 & 0.2985 & 0.7158&0.7107& 1.1 \\
 & &5 & 10 &  200& 200 & 0.1330 & 1.5399  &0.9719& 0.8922 & 21.0 \\
 \hline
 III & 6 & 5 & 10 & -200 &-200 & 0.2030 & 0.3820 &0.3919 &0.3833& 1.4 \\
 & &5 & 10 &  200& 200 & 0.0940& 10.3451  &0.9870& 0.1873 & 290.0 \\
 \hline
  IV & 8 & 5 & 10 & -200 &-200 & 0.1557 & 0.3969 &0.2532& 0.2432& 1.6 \\
 & &5 & 10 &  200& 200 & 0.0690& 79.4930  &0.9813& 0.0624 & 4052.0 \\

\midrule

\midrule
Mod Case &$N$ & $\Omega_0$ & $\Omega$ & $\Delta_0$ & $\Delta$  & $|J_W|$ & $|I_W|$  &$F_0$& $F$ & $T_g$ \\

 \hline
I & 4 &1 & 10 & -201.83 &-200 & 0.0573 & 0 & 0.9899&0.9550&6.2 \\
 & &1 & 10 &  201.32& 200 & 0.0264 & 0  &0.9908& 0.9387 & 13.5 \\

 \hline

 II & 4 &5 & 10 & -201.54 &-200 & 0.2867 & 0 & 0.9914&0.9826& 1.2 \\
 & &5 & 10 &  200.56& 200 & 0.1326 & 0  &0.9912&0.9824 & 2.8 \\
 \hline
 III & 6 & 5 & 10 & -201.50 &-200 & 0.2027 & 0 &0.9356 &0.9134& 1.7 \\
 & &5 & 10 &  201.08& 200 & 0.0937& 0  &0.9692& 0.9315 & 3.8 \\
 \hline
  IV & 8 & 5 & 10 & -201.52 &-200 & 0.1555 & 0 & 0.8549&0.8144& 2.3 \\
 & &5 & 10 &  201.40& 200 & 0.0687& 0  &0.9141& 0.8389 & 5.2 \\
 
 \midrule

\midrule
Opt Case &$N$ & $\Omega_0$ & $\Omega$ & $\Delta_0$ & $\Delta$  & $|J_W|$ & $|I_W|$  &$F_0$& $F$ & $T_g$ \\

\hline
I& 4 &1.10 &9.56 &-202.77  & -201.11 & 0.0600 & 0.0095 & 0.9932& 0.9626 &6.0 \\
& &1.06 & 10.00 & 208.31 &207.03 & 0.0268& 0.0032 & 0.9953&0.9470 & 13.3 \\

\hline
II& 4 & 4.87& 10.17 & -215.64 & -215.35 & 0.2794 & 0.0427 &0.9924 & 0.9865 & 1.2 \\
& &5.01 & 9.99 & 198.72 & 198.14 & 0.1344 & 0.0193  & 0.9948& 0.9844 & 2.8\\

\hline
III&6&4.50& 11.00 & -216.36 & -215.84 & 0.2136 & 0.0034 &0.9832 & 0.9629 & 1.4\\
& &4.50 & 9.52& 209.71  &208.76 &0.0759 &0.0016   &0.9809& 0.9448 & 4.5 \\
\hline
  IV&8& 4.51 & 10.61 & -208.90 & -207.17 & 0.1568 & 0.0478  &0.9725&0.9238&2.2 \\

& &4.51 &10.92 &  212.20 & 210.89&0.0702 & 0.0312  &0.9404 &0.8617 &5.1  \\
\hline
\end{tabular}
}

\end{table}

 In order to balance the impact of $J_W$ and $I_W$ both of which contribute a high-quality excitation transfer based on the effective three-level system, a precise control of the global parameters $(\Omega_{0},\Omega,\Delta_0,\Delta)$ is of great importance. Here, we employ the classical Genetic Algorithm (GA) as used in our recent work \textcolor{black}{\cite{PhysRevApplied.17.024014}}, focusing on maximizing the transfer fidelity $F$ in the presence of dephasing which serves as the cost function. \textcolor{black}{GA is a kind of random search algorithm inspired by Darwin's theory of natural selection and evolution, which usually contains the population initialization, fitness assessment, selection, crossover and mutation. GA starts optimization from a series of temporary results and iterates them simultaneously, which allows it to avoid local maximum and realize parallel computation easily. Thus, it offers a global optimization within the given search region for all parameters. The aim of GA's usage in our work is to find out a set of parameters to make the final transfer efficiency have the best fitness. }For implementing a fine and efficient tuning, the required ratio of $(\Omega_0/\Omega$, $\Delta_0/\Delta)$ is preset before optimization treating as the non-optimal parameters initially. Then, we add a small variation of $\pm 10\%$ with respect to every non-optimized value e.g. $\Omega_0\to (0.9,1.1)\Omega_0$ serving as the search region
 and perform the global optimization, while preserving the ratio almost unvaried.

 \begin{figure}
\centering
\includegraphics[width=0.6\textwidth]{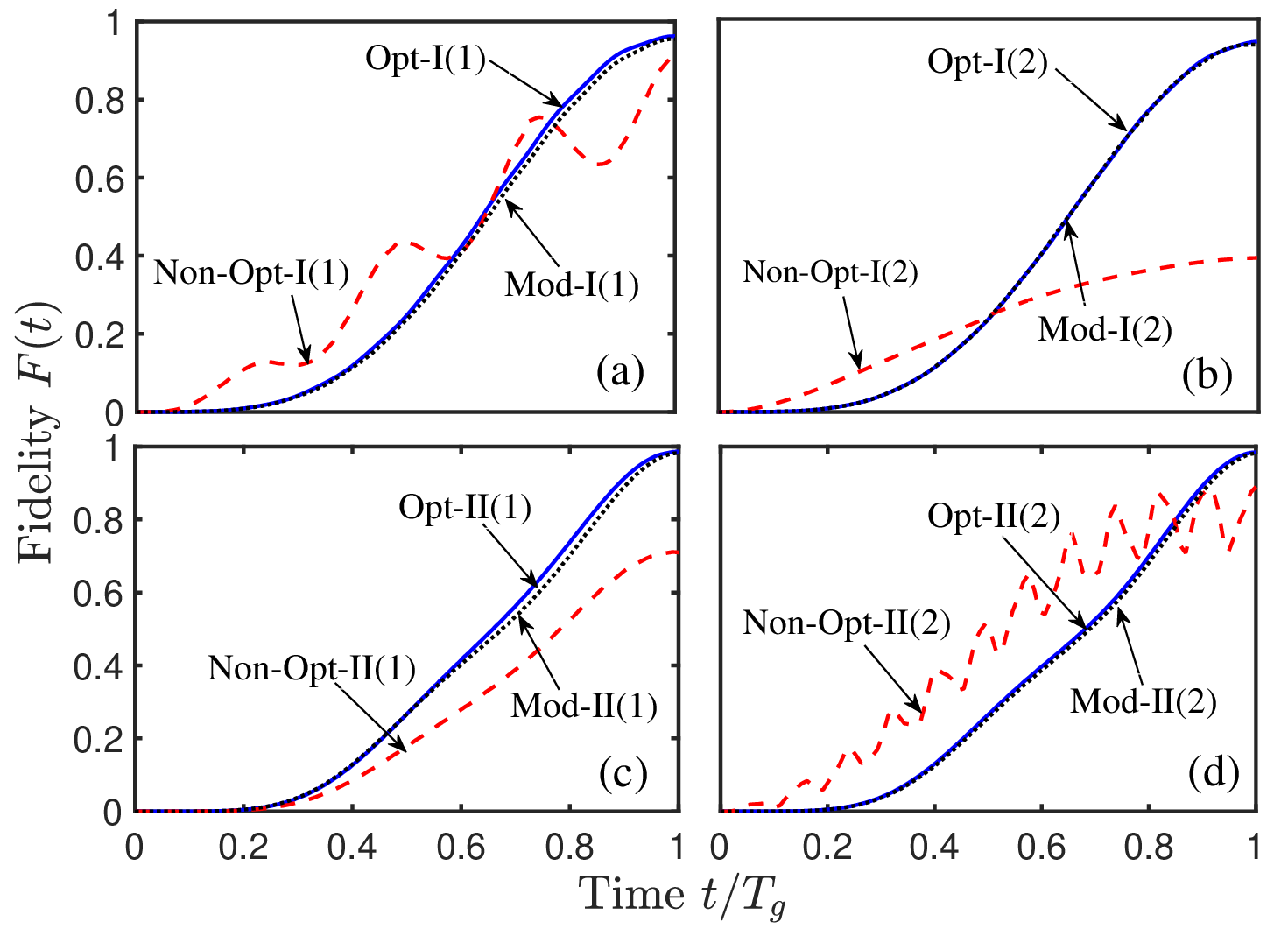}
\caption{The realistic fidelity $F(t)$ of the state initially evolving from $|\Phi_1(0)\rangle$ for $N=4$ where different ratios of Rabi frequency are used: (a-b) $\Omega_0/\Omega = 1/10$ and (c-d) $\Omega_0/\Omega = 5/10$. Here, the Non-Opt, the Mod and the Opt cases are comparably presented. Numbers 1 or 2 in parentheses stand for the case using negative or positive detunings.
All parameters are given in Table \ref{table1:pulse} except for $\Gamma/2\pi = 0.1$ MHz.}
\label{fig3:yh}
\end{figure}

 Table \ref{table1:pulse} shows the optimized parameters and the obtained fidelities for the $N$-atom excitation transport, where the effective coupling strength $J_W$ and the on-site potential $I_W$ are explicitly calculated. For a fair comparison, the Non-Opt and the Mod cases are separately presented. Remember here the value of $F_0-F$ represents the pure dephasing error which is also inversely proportional to the evolution time $T_g$. Firstly, it is obvious that the Non-Opt Cases suffer from a relatively lower fidelity in practice as plotted in Fig.\ref{fig3:yh}(a-d) by the red-dashed lines for $N=4$, because a finite on-site potential $|I_W|$ affects. In particular, if $|I_W|\gg|J_W|$ (see Fig.\ref{fig3:yh}b with a positive detuning) the evolution time becomes severely
  elongated due to the far off-resonant coupling of intermediate state $|\Phi_m\rangle$. As a consequence, the transfer process suffers from a large dephasing error and the final fidelity can be lower than 0.40. We note that, as $N$ increases this value turns to be even smaller. For example, it will need $T_g\approx 290$ $\mu$s to ensure the adiabatic transfer for $N=6$; however, the fidelity has been lowered to 0.1873 due to the significant dephasing error ($F_0-F\approx 0.8$). For $N=8$ and a positive detuning, the evolution time has been elongated to be larger than 4000 $\mu s$ with the achievable fidelity far below 0.1.

   Fortunately, we find the simplest way to solve this difficulty is precisely tuning the detuning value, e.g. $\Delta_0$, allowing the on-site potential to be $|I_W|=0$ as done by the Mod Case. When we only slightly modify $\Delta_0$ (keep other parameters unvaried) the practical fidelity denoted as $F$ which is irrespective of detuning signs, can reveal a dramatic increase alongside with a significant reduction in the evolution time. As shown in Fig.\ref{fig3:yh} labeled by the black-dotted lines, we find the behavior of dynamic evolution becomes smooth and the fidelity quickly reaches a high value (close to 1.0) which apparently outperforms the Non-Opt protocol. However, from Table \ref{table1:pulse} we see if $N$ grows a high fidelity can not be preserved by simply modifying one parameter of the detuning.

 In order to reach a higher-fidelity state transfer that is significantly robust to the intrinsic dephasing error, especially for a larger $N$ case, we implement the transfer procedure with globally optimal parameters, aiming at minimizing the infidelity $1-F$ caused by the dephasing with a shorter evolution time \textcolor{black}{\cite{PhysRevA.94.063411}}. The results of Opt-Case I and II are displayed in Fig.\ref{fig3:yh}(a-d) by the blue-solid lines and the corresponding optimized parameters are reported in Table \ref{table1:pulse}. With global optimization in which all parameters $(\Omega_0,\Omega,\Delta_0,\Delta)$ are varied by $\pm 10\%$ around its pre-set ratio $\Omega_0/\Omega = 1/10$ for Case I and $\Omega_0/\Omega = 5/10$ for Case II we observe a clear improvement (especially for $N=6$ and $8$) in $F$ outperforming the Non-Opt and Mod cases. Because, when the optimized parameters are precisely tuned, it could lead to a larger coupling $|J_W|$ although at the cost of a non-zero but tiny $I_W$ value, which provides a higher fidelity for any detuning. Apparently, a higher fidelity is mainly contributed by a stronger coupling strength alongside with the reduction in the total evolution time. 
 According to the definition of $J_W$, a larger $|J_W|$ depends on increasing the absolute laser Rabi frequencies, because $|J_W|\propto \Omega_0\Omega$. So the Opt Case II supports the best fidelity above 0.98, which manifests that our scheme can show a high-fidelity state transfer against the systematic dephasing error in practice.

  Furthermore, we explore the Opt Case for $N=6$ and $8$. From Table \ref{table1:pulse}, if we use the best case of $\Omega_0/\Omega = 5/10$ to transfer the state, the effective couplings strength $|J_W|$ still decreases with the growth of $N$ consequently requiring more times. In fact, we find that to implement the excitation transport, the fidelity inevitably decreases with $N$ albeit a global optimization has been applied \textcolor{black}{\cite{Eur.Phys.J.D.55.711-721}}. Because the transfer within the pure $Z$ subspace can not be perfectly preserved when more atoms are involved (the original system contains more intermediate states).
  However, as compared with the Non-Opt and Mod Cases for same conditions, i.e. $N=6$ and $8$, our scheme still holds sufficient robustness against the systematic dephasing errors by keeping the final fidelity at a higher level.

\section{Error-tolerant excitation transport}\label{sec4}

\subsection{Error tolerance in the transport}

\begin{figure}
\centering
\includegraphics[width=0.6\textwidth]{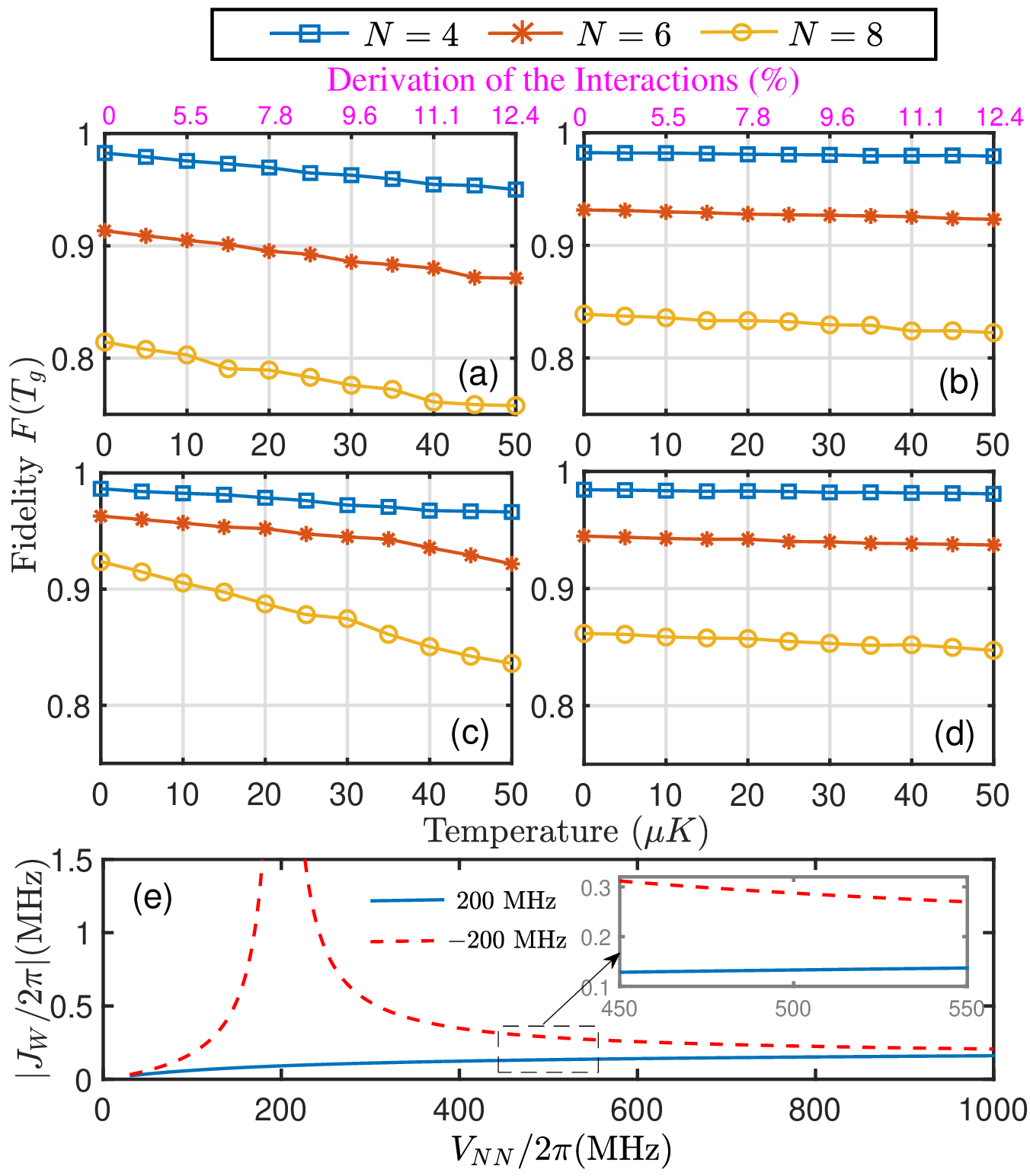}
\caption{Dependence of the transfer fidelity $F(t)$ at $t=T_g$ on the deviation of interactions (top axis) which is induced by finite atomic temperature $T\in [0,50]$ $\mu$K (bottom axis). Every point is obtained by averaging over 100 realizations of uncertain atomic positions in the full model. 
(a-b) correspond to the cases of Mod II-IV with a negative and positive detuning. Analogously, (c-d) are for the Opt II-IV cases. The fidelity number $F$ at $T=0$ is the same as given by Table I. We note that the numerical results presented are also taking account of a fluctuated NNN interaction. (e) The spin-exchange coupling strength $J_W$ as a function of the NN interaction $V_{NN}$ for $\Omega_0/\Omega = 5/10$,  $\Delta_0=\Delta = \pm 2\pi\times 200 $ MHz and $N=4$. Inset illustrates an amplified plot for the change of $J_W$ within the range of interaction deviation.
}
\label{fig4:error}
\end{figure}

Up to now, we have shown that the transfer of single-excitation state between two marginal array atoms
can be efficiently achieved by using the optimal control of minimal parameters (the Opt case), which also displays good robustness to the systematic dephasing error. \textcolor{black}{In fact, the array atoms can not be exactly frozen even for a sufficiently low temperature $T$ in experiment \textcolor{black}{\cite{Nat.Phys.17.1324}}, thus it is more instructive to explore the robustness of our scheme against other technical errors.} Because of the finite temperature the atomic positions are randomly spread leading to the fluctuated $V_{NN}$ and $V_{NNN}$ that affect excitation transfer between the atoms \textcolor{black}{\cite{New.J.Phys.19.123015,PhysRevApplied.13.024008}}. Here, we express the interatomic distance as $r_{jk}^\prime = |r_{jk}+\delta r_j-\delta r_k| $, correspondingly, the vdWs interaction between atoms turns to be $V_{jk}^\prime = C_6/(r_{jk}^\prime)^6$. Then the Hamiltonian (\ref{ham}) of the system is rewritten as
\begin{eqnarray}
    H &=& \sum_{j}\Omega_j \sigma_x^j+ \sum_j \Delta_j\sigma_{rr}^j
+\sum_{j<k} V_{jk}^\prime \sigma_{rr}^j\sigma_{rr}^{k} 
    \label{ham1}
\end{eqnarray}
The uncertainty $\delta r_j,\delta r_k$ in atomic positions results in a fluctuated vdWs interaction $V_{jk}^\prime$, which consists of
\begin{equation}
V_{jk}^\prime\approx V_{jk}-\frac{6C_6}{|r_{jk}|^7}(\delta r_{j}-\delta r_k)
\end{equation}
Since the interaction decreases rapidly as increasing the distance so we have considered the NN and NNN interactions and ignore all terms $V_{j,k>j+2}$. As a result, we require the fluctuated interactions defined by
\begin{eqnarray}
    V_{NN}^\prime = V_{NN} - \frac{6C_6}{r^7}(\delta r_{j}-\delta r_{j+1}) \nonumber\\
    V_{NNN}^\prime = V_{NNN} - \frac{6C_6}{(2r)^7}(\delta r_{j}-\delta r_{j+2})
\end{eqnarray}
with which the practical Hamiltonian under atomic position fluctuations is readily given by
\begin{equation}
    H = \sum_{j}\Omega_j \sigma_x^j+ \sum_j \Delta_j\sigma_{rr}^j
+\sum_{j} V_{NN}^\prime \sigma_{rr}^j\sigma_{rr}^{j+1} 
    +\sum_{j} V_{NNN}^\prime \sigma_{rr}^j\sigma_{rr}^{j+2}
    \label{ham1}
\end{equation}

 At a finite temperature $T$, the distribution of atomic positions in the array is approximately a Gaussian function with standard deviation $\sigma=\sqrt{k_B T/(m w^2)}$, where $\omega$ is the trap frequency \textcolor{black}{\cite{ Nat.Photon.16.724}}. Using $^{87}$Rb atoms held at a temperature of $T=50$ $\mu$K in a trap with frequency $\omega = 2\pi\times 147$ kHz, the maximal uncertainty is calculated to be $\sigma = 0.0775$ $\mu$m (corresponding to the interaction deviation of $|(V_{NN}^\prime-V_{NN})/V_{NN}|=6\sigma/r\approx 12.4\%$) which is still much smaller than the NN distance $r = 3.76$ $\mu$m as estimated in our paper. In the calculation the real position deviation $\delta r_j$ of every atom should be extracted from the 1D Gaussian distribution that contributes intrinsic randomness for each measurement. The final results depend on a sufficient sampling of atomic positions, in analogy with the fluctuated interactions.

 From the view of effective model under the WC condition (see Eq.\ref{wccn}) the role of interactions $V_{NN}$ and $V_{NNN}$ has been essentially correlated with the effective coupling $J_W$ and on-site potential $I_W$ that affects the excitation transport. In principle, if $J_W$ and $I_W$ are insensitive to the variation of interactions our scheme can have good robustness against the position error. Such a scenario is quite different from the antiblockade facilitation protocols \textcolor{black}{\cite{PhysRevLett.125.133602,PhysRevLett.128.013603}} which rely on an absolute facilitation condition $\Delta = -V_{NN}$ \cite{PhysRevLett.98.023002} to achieve the transport, which should be extremely affected by the atomic position fluctuation \textcolor{black}{\cite{Phys.Rev.A.99.060101}}.

To investigate the robustness of our scheme, especially against the dominant position error between array atoms, in Fig.\ref{fig4:error} we calculate the average fidelity after considering 100 random realizations for a finite atomic temperature $T\in [0,50]$ $\mu$K, where the corresponding deviation of interaction strength is also given (see top axis). Apparently, both the Mod and Opt cases [see Fig.\ref{fig4:error}(b) and (d)] can hold a better robustness to the position error in practice if positive detunings are applied. E.g. in the Opt case II with $N=4$ the final fidelity keeps around 0.9809 at $T=50$ $\mu$K which means a very tiny reduction of 0.0035 impacted by the position deviation. Even for $N=8$ in Opt case IV we have a small reduction of 0.0145 in the transfer fidelity when the relative deviation of interaction is about $\sim 12.4 \%$ which represents the robustness and high-fidelity of scheme can be preserved in a larger system as long as the detuning is positive. In contrast, for a negative detuning the dynamic behavior of excitation transport is more sensitive to the thermal fluctuation of atomic positions, see Fig.\ref{fig4:error}(a) and (c), apparently decreasing the transport efficiency.

This distinct result due to different signs of detunings can be easily understood by the change of $|J_W|$ due to the interaction fluctuation. As shown in Fig.\ref{fig4:error}(e), the effective coupling strength $|J_W|$ versus different NN interactions is displayed. It exhibits that, if the detuning is negative labeled by the red-dashed line, there exists a special singularity at which $\Delta_0+V_{NN} = 0$ 
providing the regime of excitation facilitation (avoided by our scheme) \textcolor{black}{\cite{PhysRevA.90.011603}}. 
Beyond that, $|J_{W}|$ dramatically decreases to be plateau at \textcolor{black}{$\Omega_0\Omega/|\Delta_0|$} if $V_{NN}\to \infty$.
As for a positive detuning labeled by the blue-solid line, the $J_W$ function features an apparent soft-core property that varies more slowly with the change of $V_{NN}$, which can make the excitation dynamics insensitive against the atomic position error, promising for an error-tolerant excitation transfer \textcolor{black}{\cite{Phys.Rev.X.14.011025}}. In the inset of Fig.\ref{fig4:error}(e) we amplify the plot in the regime where the disorder-induced interaction fluctuation is about $\pm 10 \%$. We note that the effective coupling strength $|J_W|$ for a position detuning keeps almost unvaried with respect to the fluctuated interaction perfectly agreeing with our finding that the transport fidelity remains high under a large position fluctuation in the positive-detuning case.

\begin{figure}
\centering
\includegraphics[width=0.7\textwidth]{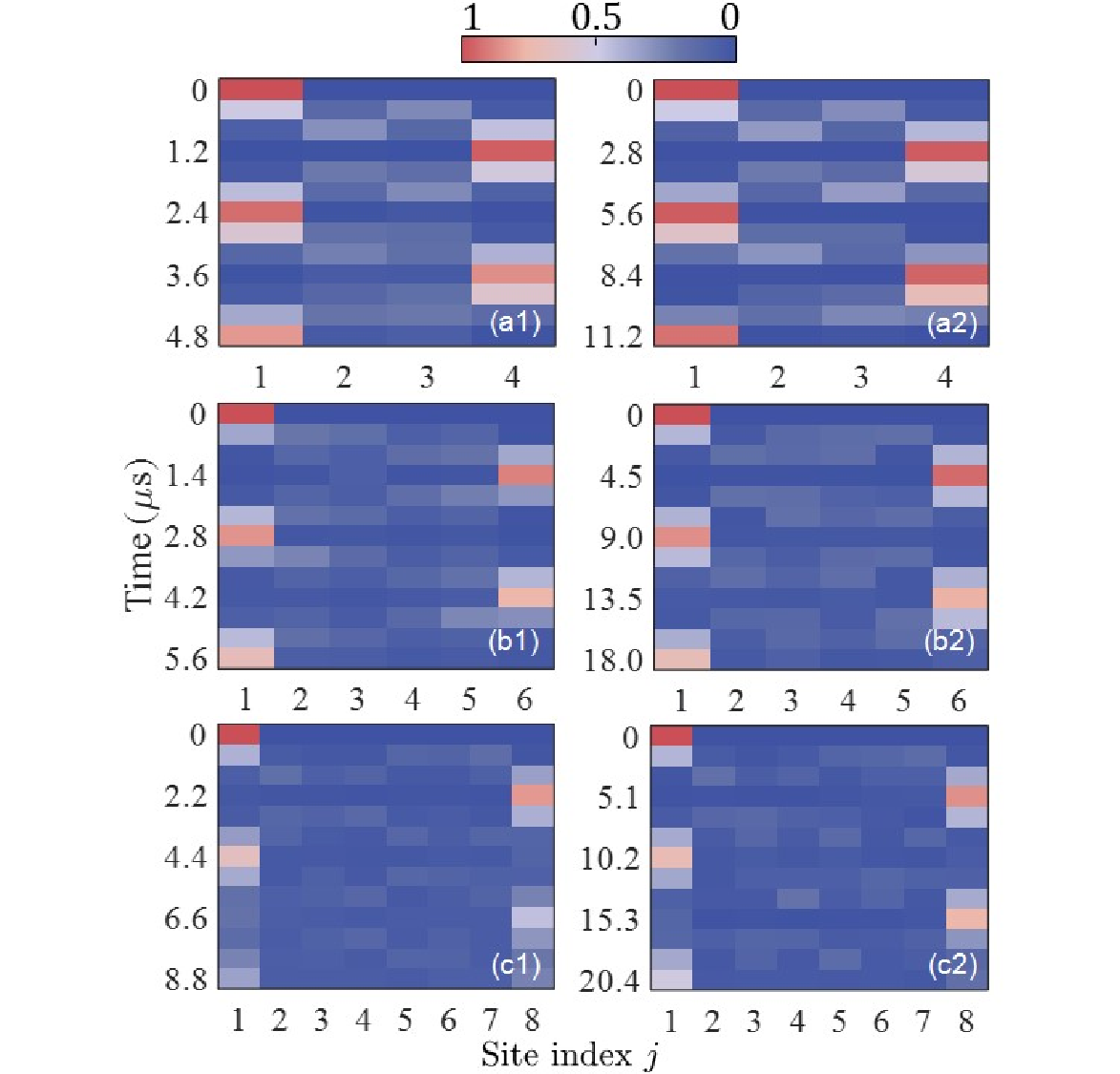}
\caption{From top to bottom: realistic long-time dynamics of single Rydberg excitation in the presence of dephasing and decay errors, which are obtained by averaging over 100 realizations with atomic position fluctuations for $T = 50$ $\mu$K, corresponding to the Opt Cases II-IV in Table \ref{table1:pulse}. Left and right panels represent the cases of negative and positive detunings, respectively. 
Here, the positive-detuning cases feature higher fidelity and higher robustness for a long-time propagation.
}
\label{fig5:long}
\end{figure}

\subsection{Long-time excitation transport between two marginal atoms}

We have derived an effective model in the limit of WC condition based on which a quantum-preserved transport of Rydberg excitations having both high-fidelity and robustness features, is optimally achieved. This can preserve the excitation of atoms within the $Z$ subspace and meanwhile strongly reduce the intermediate-state populations by bridging two marginal atoms in a simple three-level system. Here we discuss whether the high-quality transport could be persistently performed for any direction in many-atom arrays \textcolor{black}{\cite{PhysRevLett.132.223201}}. \textcolor{black}{In order to explore the realistic excitation transport for a long-time duration between marginal atoms we simulate the population dynamics under the full Hamiltonian (\ref{ham}) for different signs of detuning, corresponding to the Opt Cases II-IV of Table I. From the experimental point of view, a long-time transport might not be well isolated from the dissipation caused by spontaneous decay. Therefore, we use $\gamma/2\pi = 0.4$ kHz denoting the decay rate from state $|73S_{1/2}\rangle$ and $\Gamma/2\pi=0.1$ MHz throughout. Also, the simulation is carried out at a finite temperature $T=50$ $\mu$K arising a 12.4$\%$ deviation for the NN and NNN interactions.}

As shown in Fig.\ref{fig5:long}, the error-tolerant dynamics with a positive detuning (right panels) can obtain a clear improvement as compared to the case with a negative detuning (left panels), even in a long-time excitation transport. For $N=8$ the transfer fidelity at $t=4T_g$ can sustain \textcolor{black}{$F=0.5302$} [see (c2)] which means the excitation probability can largely return to state $|\Phi_1\rangle$ after two rounds of propagation; whilst this value is lowered to 0.3651 [see (c1)] when the detuning is negative.

\section{Conclusion}\label{sec13}

We develop a realistic scheme for transporting excitation in a 1D Rydberg-atom array with high-fidelity and high-robustness, which utilizes the effective spin-exchange interaction induced by the diagonal vdWs potential \cite{PhysRevLett.123.063001}. With the optimal control strategy, we show that, the excitation transport can persistently stay in the single-excitation Zeno subspace ensuring a strong tolerance to both the intrinsic dephasing error as well as the atomic position fluctuation.
Under the WC condition, this scheme only relies on precisely modulating two marginally atomic couplings while the controls exerted on all intermediate atoms are globally engineered, which can significantly reduce the experimental difficulty by avoiding local controls as in a PST protocol \textcolor{black}{\cite{PhysRevA.72.034303}}. We furthermore present that there may achieve a long-time and high-quality transport between two marginal atoms for a multi-atom array \textcolor{black}{\cite{PhysRevLett.98.230503}}. Our scheme is feasible in experimental operations which can contribute a newly-efficient and stable method for long-distance quantum information transfer, and even for the construction of quantum networks \textcolor{black}{\cite{Nature.558.264,Laser.Photon.Rev.16.2100219,science.362.6412}} .

\begin{appendices}

\section{The effective weak-coupling model}\label{secA1}

We consider a 1D array of individually trapped cold atoms, dressed by laser fields $\Omega_j$ that couple the ground state $|g_j\rangle$ to the Rydberg state $|r_j\rangle$ with a site-dependent detuning $\Delta_j$. Two excited atoms will interact via a diagonal vdWs potential $V_{jk}=C_6/{r}_{jk}^6$ with ${r}_{jk}=|\bold{r}_j-\bold{r}_k|$ the distance.
Recently, Yang {\it et. al.} \textcolor{black}{\cite{PhysRevLett.123.063001}} derived an effective Hamiltonian for a single Rydberg excitation by applying the second-order Van Vleck perturbation theory \textcolor{black}{\cite{Journal1980}}, given by
\begin{eqnarray}
     H_{eff,N} &=& \sum_{j=1}^{N-1} J_{j,j+1} (|\Phi_{j+1}\rangle\langle \Phi_j|+|\Phi_{j}\rangle\langle \Phi_{j+1}|)+ \sum_{j=1}^{N-2} J_{j,j+2} (|\Phi_{j+2}\rangle\langle \Phi_j|+|\Phi_{j}\rangle\langle \Phi_{j+2}|) \nonumber \\
    &+&\sum_{j=1}^N I_j|\Phi_j\rangle\langle \Phi_j|
    \label{eff1}
\end{eqnarray}
with the off-resonant condition of $|\Delta_j|,|\Delta_j+V_{jk}|\gg\Omega_j$,
where the effective NN and NNN spin-exchange interactions are 
\begin{equation}
   J_{j,j+1} = \sum_{k=j,j+1}\frac{\Omega_j\Omega_{j+1}V_{NN}}{2\Delta_{k}(\Delta_{k}+V_{NN})},J_{j,j+2} =\sum_{k=j,j+2}\frac{\Omega_j\Omega_{j+2}V_{NNN}}{2\Delta_{k}(\Delta_{k}+V_{NNN})}
\end{equation}
and the Ising-type on-site potentials take the following form as
\begin{equation}
   I_j = \Delta_j+\frac{2\Omega_j^2}{\Delta_j}+\sum_{k =j\pm1}\frac{\Omega_{k}^2V_{NN}}{\Delta_{k}(\Delta_{k}+V_{NN})}+\sum_{k = j\pm 2}\frac{\Omega_{k}^2 V_{NNN}}{\Delta_{k}(\Delta_{k}+V_{NNN})}
\end{equation}
Here, $|\Phi_j\rangle = |g_1g_2...r_j...g_N\rangle$ with $j = 1,2,...,N$ represents the collective singly-excited state and $\Pi_1 = \sum_j|\Phi_j\rangle\langle \Phi_j|$ forms a quasi-degenerate one-excitation subspace. The resulting effective Hamiltonian $H_{eff,N}$ denotes that the dynamical transport involving one Rydberg excitation can be restricted inside the subspace $\Pi_1$. For simplicity, we have chosen the vdWs potential $V_{jk}$ up to $V_{NNN}$ with 
$k= j\pm2$, and the NN interaction is described by $V_{NN}$ for $k=j\pm1$.


In analogy with other dressing protocols \textcolor{black}{\cite{PhysRevLett.116.230503,NewJ.Phys.13.073044,PhysRevA.108.043308}} , the goal here is to achieve a high-quality excitation transport between two marginal array atoms. For systems dominated by the NN interaction,
we can engineer the strength of different spin-exchange interaction $J_{j,j+1}$ by tuning local parameters $\Omega_j$ and $\Delta_j$ to be a PST condition $J_{j,j+1}=J\sqrt{j(N-j)}$ alongside with same on-site energy $I = I_j$ \textcolor{black}{\cite{nat.comms.7.11339}}. However, a fully local control of all pulse parameters with the increasing number of atoms, makes the protocol intrinsically challenging in practice. Thereby our target lies in how to achieve the transport with minimally global controls.

Next we consider the system involves two marginal atomic qubits with a relatively weak coupling, namely, $\Omega_0 \ll \Omega$ and arbitrary detunings $\Delta_0,\Delta$, serving as the WC condition \textcolor{black}{\cite{PhysRevLett.99.060401}}. Note that $\Omega_{0}=\Omega_{1,N}$, $\Omega = \Omega_{2,...,N-1}$ and same for $\Delta_0,\Delta$ which means at most four parameters $(\Omega_0,\Omega,\Delta_0,\Delta)$ have to be determined even if $N$ is sufficient. Also the facilitation regime with $V_{NN},V_{NNN}$ compensated by the detunings should be avoided, since a small fluctuation of atomic positions can hinder the excitation transport in this regime \textcolor{black}{\cite{Phys.Rev.Lett.118.063606}}.
Therefore, incorporating with the WC assumption, the NN interaction is rewritten as
\begin{equation}
    J_{0} =J_{12} =J_{N-1,N}=\frac{\Omega_0\Omega V_{NN}}{2\Delta_0(\Delta_0+V_{NN})}+\frac{\Omega_0\Omega V_{NN}}{2\Delta(\Delta+V_{NN})}, \text{others } J= \frac{\Omega^2 V_{NN}}{\Delta(\Delta+V_{NN})} \label{nni}
\end{equation}
satisfying $J\gg J_0$ (because $\Omega\gg\Omega_0$) and the NNN interaction is
\begin{equation}
    J_0^\prime=J_{13} =J_{N-2,N}= \frac{\Omega_0\Omega V_{NNN}}{2\Delta_0(\Delta_0 + V_{NNN})}+\frac{\Omega_0\Omega V_{NNN}}{2\Delta(\Delta + V_{NNN})}, \text{others } J^\prime = \frac{\Omega^2 V_{NNN}}{\Delta(\Delta + V_{NNN})} \label{nnni}
\end{equation}
With the above definitions (\ref{nni}) and (\ref{nnni}), the effective Hamiltonian $H_{eff,N}$ can be obtained as
\begin{eqnarray}
    H_{eff,N} &=& J_0(|\Phi_{2}\rangle\langle \Phi_1|+|\Phi_{1}\rangle\langle \Phi_{2}|)+J_0(|\Phi_{N}\rangle\langle \Phi_{N-1}|+|\Phi_{N-1}\rangle\langle \Phi_{N}|)  \nonumber \\
    &+& J_0^\prime(|\Phi_{3}\rangle\langle \Phi_1|+|\Phi_{1}\rangle\langle \Phi_{3}|)+J_0^\prime(|\Phi_{N}\rangle\langle \Phi_{N-2}|+|\Phi_{N-2}\rangle\langle \Phi_{N}|) \nonumber\\
    &+& J\sum_{j=2}^{N-2}(|\Phi_{j+1}\rangle\langle \Phi_j|+|\Phi_{j}\rangle\langle \Phi_{j+1}|) +J^\prime\sum_{j=2}^{N-3}(|\Phi_{j+2}\rangle\langle \Phi_j|+|\Phi_{j}\rangle\langle \Phi_{j+2}|)  \nonumber\\
    &+& \sum_{j=1}^N I_j|\Phi_j\rangle\langle \Phi_j|
    \label{effwc}
\end{eqnarray}

To further evaluate the transport quality based on such a WC assumption, we focus on our interest in the subspace $\Pi_1$ formed by $N$ singly-excited states, by which the Hamiltonian (\ref{effwc}) can be exactly rewritten in a matrix form as
\begin{equation}
    H_{eff,N}=H_0 + H_a
    \label{d}
\end{equation}
where the first term $H_0$ is dominant due to the strong spin-exchange coupling $J\gg J_0$
\begin{equation}
H_0=J\left[\begin{array}{ccccccc}
     2&0&0&\cdots&0&0&0  \\
     0&1&1&\cdots&0&0&0   \\
     0&1&0&\cdots&0&0&0 \\
    \vdots&\vdots&\vdots&\ddots&\vdots&\vdots&\vdots  \\
    0&0&0&\cdots&0&1&0  \\
    0&0&0&\cdots&1&1&0  \\
    0&0&0&\cdots&0&0&2
\end{array}\right]_{N\times N}
\label{H0}
\end{equation}
and the secondly auxiliary Hamiltonian is
\begin{equation}
H_a=\left[\begin{array}{ccccccccc}
     I_1-2J&J_{0}&J_{0}^\prime&0&\cdots&0&0&0 &0 \\
     J_{0}&I_2-J&0&J^\prime&\cdots&0&0&0&0   \\
     J_{0}^\prime&0&I_3&0&\cdots&0&0&0&0 \\
     0&J^\prime&0&I_4&\cdots&0&0&0&0  \\
    \vdots&\vdots&\vdots&\vdots&\ddots&\vdots&\vdots&\vdots&\vdots  \\
    0&0&0&0&\cdots&I_{N-3}&0&J^\prime&0  \\
    0&0&0&0&\cdots&0&I_{N-2}&0&J_{0}^\prime\\
    0&0&0&0&\cdots&J^\prime&0&I_{N-1}-J&J_{0}  \\
    0&0&0&0&\cdots&0&J_{0}^\prime&J_{0}&I_N-2J
\end{array}\right]_{N\times N}
\label{aa}
\end{equation}

\begin{figure}
\centering
\includegraphics[width=0.9\textwidth]{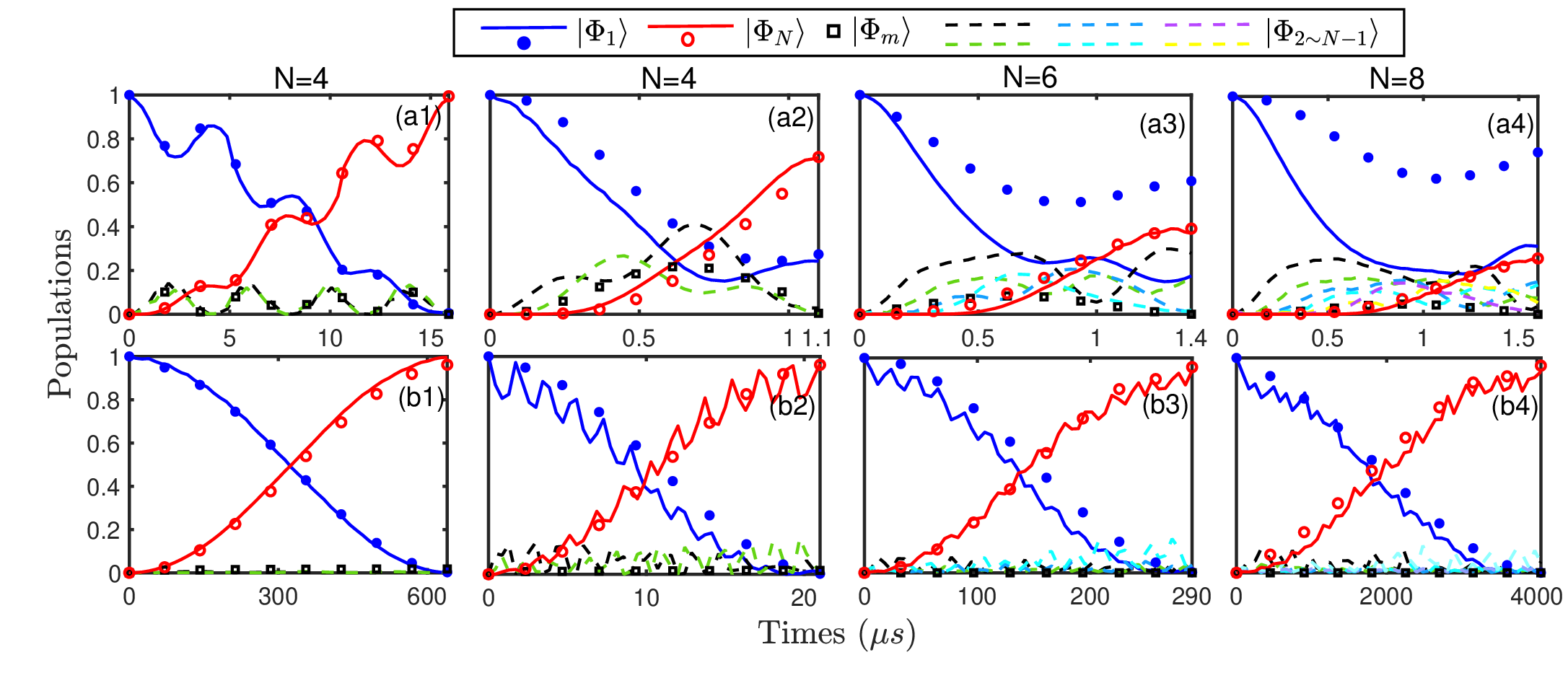}
\caption{Numerical verification of the effective WC model representing a good consistence with the rigorous dynamics. The rigorous dynamics displayed by the solid and dashed lines 
is based on solving the exact model with the full Hamiltonian $H$ in (\ref{ham}) and the effective dynamics of states $\{|\Phi_1\rangle,|\Phi_m\rangle,|\Phi_N\rangle\}$ utilizing the Hamiltonian $H_{WC,N}$ in (\ref{wcc}) is denoted by stars, squares and circles. From left to right, panels correspond to the Non-Opt Cases I-IV in Table \ref{table1:pulse}, where the upper panels are given for $(\Delta_0,\Delta)/2\pi = -200$ MHz and the lower ones for $(\Delta_0,\Delta)/2\pi = 200$ MHz. Here, no decay is considered.
}
\label{figA:dynamic}
\end{figure}

In deducing (\ref{H0}) we have shifted the energy of states $|\Phi_1\rangle,|\Phi_N\rangle$ by $2J$ for a same dimension, while, which are also subtracted in $H_a$ [see (\ref{aa})]. Thus based on $H_0$, it is obvious that $|\Phi_1\rangle$ and $|\Phi_N\rangle$ are two eigenstates with the same eigenvalue $2J$. Other eigenvalues of $H_0$ can be resolved via a reduced matrix $M$ 
\begin{equation}
M=J\left[\begin{array}{ccccc}
     1&1&\cdots&0&0  \\
     1&0&\cdots&0&0   \\
     0&0&\cdots&0&0 \\
    \vdots&\vdots&\ddots&\vdots&\vdots  \\
    0&0&\cdots&0&0  \\
    0&0&\cdots&0&1  \\
    0&0&\cdots&1&1
\end{array}\right]_{(N-2)\times (N-2)}
\end{equation}
whose eigenvalue is \textcolor{black}{\cite{New.J.Phys.13.123006}}
\begin{equation}
\epsilon_p=2J\cos[\frac{(p-1)\pi}{N-2}],p=1,2,\cdots,N-2
\end{equation}
When $p=1$, we get $\epsilon_1=2J$, which means other than eigenstates $|\Phi_1\rangle$ and $|\Phi_N\rangle$, there also exists a third eigenstate denoted as $|\Phi_m\rangle$ having same energy $2J$. By solving the eigen-equation
\begin{equation}
    H_0|\Phi_m\rangle=2J|\Phi_m\rangle 
    \label{eigeq}
\end{equation}
whose explicit eigenstate is $|\Phi_m\rangle=\frac{1}{\sqrt{N-2}}(|2\rangle+|3\rangle+\cdots+|N-1\rangle)$.
According to quantum Zeno effect, the evolution of system could remain in a Zeno subspace with inhibited quantum state transitions, \textcolor{black}{as long as the frequent measurement ($H_0$) on the system could be devised with multidimensional projections} \textcolor{black}{\cite{Phys.Rev.A.95.042132,Sci.Rep.5.11509,Nat.Phys.10.715,PhysRevLett.105.213601}}. Therefore, due to the strong coupling limit $J\gg J_0$, the whole system can be divided into $N-2$ quantum Zeno subspaces (since $H_0$ has $N-2$ different eigenvalues). Because the measurement is strong and continuous so the system will approximately evolve in the initial-state subspace formed by $Z = \{|\Phi_1\rangle,|\Phi_m\rangle,|\Phi_N\rangle\}$. After defining the projector in the $Z$ subspace by $P = \sum_{\alpha}|\alpha\rangle\langle \alpha |(|\alpha\rangle \in Z)$, we rewrite the effective Hamiltonian $H_{eff,N} $ in the matrix form under the WC condition, as
\begin{eqnarray}
H_{WC,N}&=& 2J P + PH_aP \nonumber \\
 &=&  \left[\begin{array}{ccc}
     I_1-2J&\frac{J_{0}+J_{0}^\prime}{\sqrt{N-2}}&0 \\
     \frac{J_{0}+J_{0}^\prime}{\sqrt{N-2}}&\frac{(I_2+I_3+I_4+...+I_{N-1}-2J)}{N-2}+\frac{2(N-4)J^\prime}{N-2}&\frac{J_{0}+J_{0}^\prime}{\sqrt{N-2}}   \\
     0&\frac{{J_{0}+J_{0}^\prime}}{\sqrt{N-2}}&I_N-2J \label{wcn}
\end{array}\right]+2J\mathcal{I}
\end{eqnarray}
Since the last constant term $2J\mathcal{I}$ contributes nothing to the dynamical evolution, it can be ignored. Therefore, we obtain 
\begin{eqnarray}
H_{WC,N}=  \left[\begin{array}{ccc}
     0&J_W&0 \\
     J_W&I_W&J_W   \\
     0&J_W&0 \label{wcc}
\end{array}\right]
\end{eqnarray}
where
\begin{equation}
  J_W=\frac{J_{0}+J_{0}^\prime}{\sqrt{N-2}}, I_W=\frac{\sum_{k=2}^{N-1}I_k-2J}{N-2}+\frac{2(N-4)J^\prime}{N-2}-I_1+2J \label{JIw}
\end{equation}
respectively stand for the effective spin-exchange and Ising-type interactions for the WC assumption.
That means, for systems initially in $|\Phi_1\rangle$ the excitation can be persistently transferred in the $Z$ subspace with a higher fidelity.

Let us proceed to verify the accuracy of the effective model $H_{WC,N}$ by comparing it with the rigorous dynamics solved from the Lindblad master equation in Eq.(\ref{rrh}), where no dephasing error is considered.
Figure \ref{figA:dynamic} provides a detailed comparison of the population dynamics between the full and the effective Hamiltonians representing a good consistence in general. Specifically, for a positive detuning (lower panels), the rigorous system evolves with a strong suppression of all intermediate-state populations (i.e. in $|\Phi_{2\sim N-1}\rangle$), demonstrating an adiabatic transfer between $|\Phi_1\rangle$ and $|\Phi_N\rangle$ alongside with a longer duration due to $|I_W|\gg|J_W|$, as same as analyzed by the effective model in Sec. \ref{subsec1}. However, e.g. for $N=8$ the one-time evolution requires even larger than 4000 $\mu$s [see Fig.\ref{figA:dynamic}(b4)] which makes the excitation transport scheme unavailable via a realistic Rydberg-atom system.
While turning to a negative detuning (upper panels) with which the spin-exchange ($J_W$) and Ising-type ($I_W$) interactions are competitive, the resulting dynamics of the full system becomes difficult to stay in the $Z$ subspace given by the effective model
as $N$ increases which is mainly caused by the imperfect WC condition. A discrete distribution of populations on $N-2$ intermediate states is observed leading to a lower transport fidelity.
Our target is achieving high-fidelity excitation transfer from $|\Phi_1\rangle$ to $|\Phi_N\rangle$ with strong robustness against the atomic position fluctuation. In fact, there exists a balance between the transfer fidelity and weak coupling strength \textcolor{black}{\cite{PhysRevA.98.012334}} which makes the optimal control strategy necessary, see more details in the maintext.

\textcolor{black}{\section{The effective uniform-coupling model}}

\begin{table}
\caption{\label{table:pulse}Coefficients corresponding to numerical results in Fig.\ref{figA:uniform}. All parameters including $(\Omega_0,\Omega,\Delta_0,\Delta)$ and $(J_0,J,I_1,I_2,I_3)$ are in unit of $2\pi\times$MHz. $F_0$ is the ideal transfer fidelity without $\Gamma$ and $T_g$ (in unit of $\mu$s) is the evolution time. 
 }

\renewcommand{\arraystretch}{1.3}
\setlength{\tabcolsep}{0mm}{
\begin{tabular}{c@{\hspace{10pt}}c@{\hspace{10pt}}c@{\hspace{10pt}}c@{\hspace{10pt}}c@{\hspace{10pt}}c@{\hspace{10pt}}c@{\hspace{10pt}}c@{\hspace{10pt}}c@{\hspace{10pt}}c@{\hspace{10pt}}c@{\hspace{10pt}}c@{\hspace{10pt}}c@{\hspace{10pt}}c}
\hline
Non-Opt&$N$ & $\Omega_0$ & $\Omega$ & $\Delta_0$ & $\Delta$  &$F_0$ & $T_g$&$|J_{0}|$ &$|J|$ &$|I_{1}|$ &$|I_{2}|$ &$|I_{3}|$ \\
\hline
I & 4 &10 & 10 & 200 &200  &0.5447 &1.2&0.3574 &0.3574 &201.376 & 201.734 &201.734 \\
\hline
 II & 6 &10 & 10 & 200 &200  & 0.3831 & 1.7 & 0.3574 & 0.3574 & 201.376 & 201.734 &201.752 \\
 \hline
 III & 8& 10 & 10 & 200 &200 &0.3044& 2.2&0.3574 &0.3574& 201.376 & 201.734 &201.752  \\

\midrule
 
\midrule
Opt &$N$ & $\Omega_0$ & $\Omega$ & $\Delta_0$ & $\Delta$   &$F_0$ & $T_g$ &$|J_0|$ &$|J|$ &$|I_{1}|$ &$|I_{2}|$ &$|I_{3}|$  \\
\hline
I & 4 &10 & 10 & 204.61 &204.21  &0.9582 &1.3 & 0.3475&0.3479 & 205.944 & 205.901 &205.901\\
\hline
 II & 6 &10 & 10 & 204.20 &203.79  & 0.8942& 1.8& 0.3484 & 0.3488 & 205.550 & 205.487 & 205.506\\
 \hline
 III & 8& 10 & 10 & 201.85 &201.42 &0.8305 & 2.3 & 0.3536 & 0.3541 & 203.216 & 203.144 &203.163\\
\hline

\end{tabular}
}

\end{table}

\begin{figure}
\centering
\includegraphics[width=0.6\textwidth]{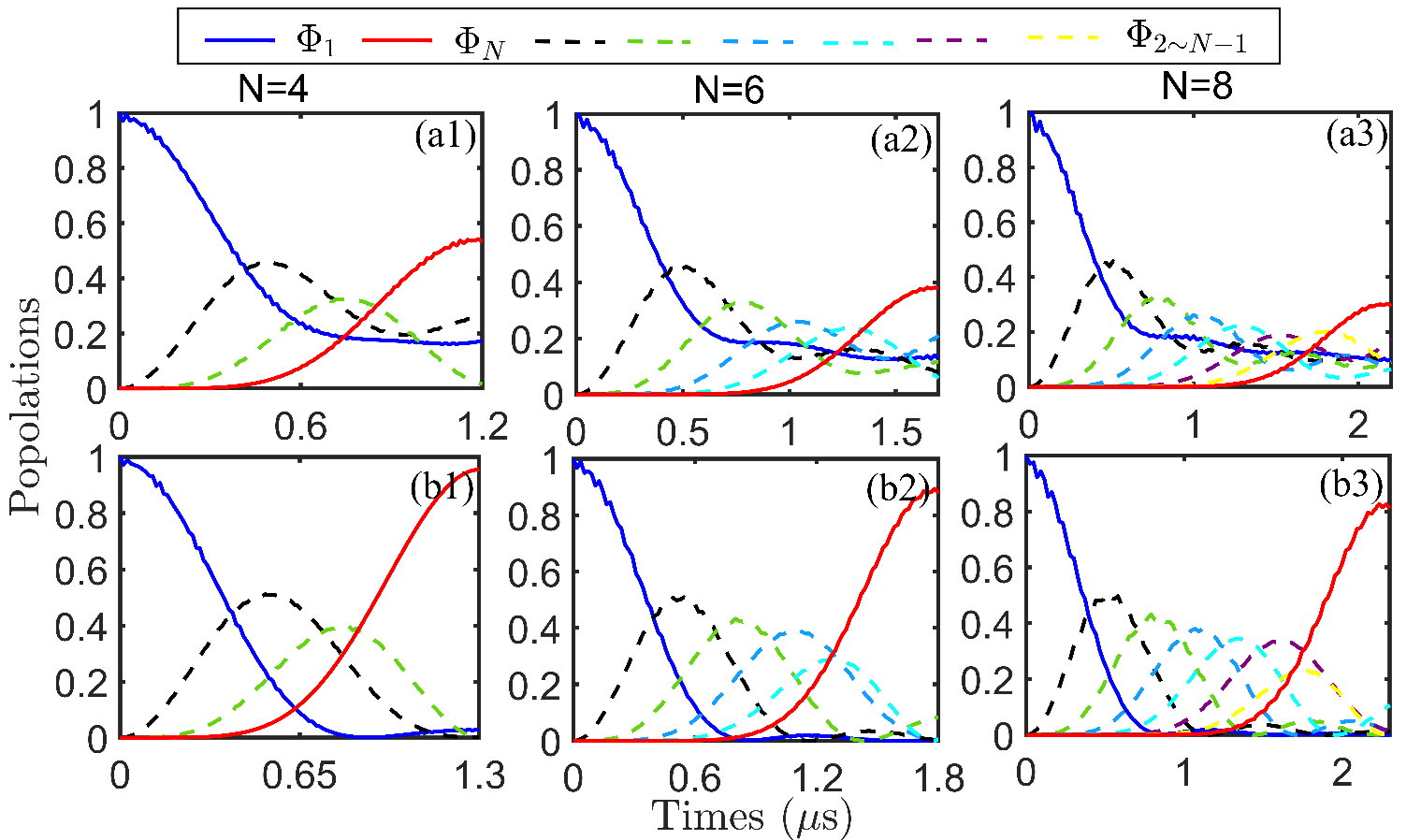}
\caption{The expected population evolution under the uniform-coupling condition in the absence of dephasing error, where $(\Omega_0,\Omega)/2\pi=10$ MHz. Panels (a1-3) correspond to the Non-Opt case where $(\Delta_0,\Delta)/2\pi=200$ MHz, and panels (b1-3) to the Opt case where $\Delta_0$ and $\Delta$ are optimal values. Relevant parameters are given in Table \ref{table:pulse}.
}
\label{figA:uniform}
\end{figure}

To illustrate the necessity of using the WC condition in the scheme we also explore an implementation for state transfer with the uniform-coupling $\Omega=\Omega_j$ (i.e. a non-ideal WC condition), for which 
the NN interaction is rewritten as
\begin{equation}
    J_{0} =J_{12} =J_{N-1,N}=\frac{\Omega^2 V_{NN}}{2\Delta_0(\Delta_0+V_{NN})}+\frac{\Omega^2 V_{NN}}{2\Delta(\Delta+V_{NN})}, \text{others } J= \frac{\Omega^2 V_{NN}}{\Delta(\Delta+V_{NN})} \label{bnni}
\end{equation}
and the NNN interaction is
\begin{equation}
    J_0^\prime=J_{13} =J_{N-2,N}= \frac{\Omega^2 V_{NNN}}{2\Delta_0(\Delta_0 + V_{NNN})}+\frac{\Omega^2 V_{NNN}}{2\Delta(\Delta + V_{NNN})}, \text{others } J^\prime = \frac{\Omega^2 V_{NNN}}{\Delta(\Delta + V_{NNN})} \label{bnnni}
\end{equation}
and the Ising-type on-site potentials take the form as
\begin{equation}
   I_j = \Delta_j+\frac{2\Omega^2}{\Delta_j}+\sum_{k =j\pm1}\frac{\Omega^2V_{NN}}{\Delta_{k}(\Delta_{k}+V_{NN})}+\sum_{k = j\pm 2}\frac{\Omega^2 V_{NNN}}{\Delta_{k}(\Delta_{k}+V_{NNN})}
\end{equation}
Accounting for the symmetry of array as well as the interactions, we note $I_1=I_N$, $I_2=I_{N-1}$, $I_3 = I_{4,...,N-2}$.

We further simulate the rigorous dynamics with the Lindblad master equation (\ref{rrh}) under the uniform-coupling condition. 
Figure \ref{figA:uniform} provides the full dynamical behavior of excitation transport in the absence of dephasing error. As expected, due to the invalid of WC condition the transport can not be restricted in the quasi-degenerate three-level frame tending to have a sequent transfer in the atomic array. While, for the Non-Opt system the transfer fidelity is quite low in general, as verified by numerical results shown in Fig.\ref{figA:uniform}(a1-3), due to the large difference among the on-site energies $|I_{1\sim 3}|$ that impede the transport. E.g. for $N=4$, we find $||I_{2,3}|-|I_1||\approx 2\pi\times0.358$ MHz, which is quite comparable to the values of $|J|,|J_0|$. Luckily, as long as the optimal detuning values $\Delta_0,\Delta$ are applied leading to $||I_{2,3}|-|I_1||\ll |J|,|J_0|$ (here $||I_{2,3}|-|I_1||\approx 2\pi\times 0.043$ MHz, about one order of magnitude smaller than $|J|,|J_0|$ in the Opt system), 
the coherent dynamics between the two marginal array of atoms can be established providing an apparent improvement in the fidelity. Fig.\ref{figA:uniform}(b1-3) shows the optimal dynamics under the uniform coupling, significantly outperforming the non-optimal cases (a1-3) by enabling a higher transfer fidelity. Thus, the optimal control method may also help to improve the quality of state transport in a uniform-coupling environment, however, the system still suffers from a relatively lower transfer efficiency there because the sequent transfer behavior among multi-intermediate states $|\Phi_{2,...,N-1}\rangle$ would add to difficulty in the practical implementation.




\end{appendices}

\backmatter

\section*{Declarations}

\bmhead{Acknowledgements} We thank Shi-Lei Su for helpful discussions.

\bmhead{Author contributions} JQ proposed the idea and supervised the project. PPL carried out the calculations. JQ and WPZ wrote the manuscript. All authors read and approved the final manuscript.

\bmhead{Funding} This work was sponsored by the Innovation Program for Quantum Science and Technology (Grant No.2021ZD0303200), 
the National Natural Science Foundation of China (Grants No.12174106, No.11474094, No.11104076), the Natural Science Foundation of Chongqing (Grant No.CSTB2024NSCQ-MSX1117), the Shanghai Science and Technology Innovation Project (No. 24LZ1400600), the National Key Research and Development Program of China under (Grant No.2016YFA0302001),  the Shanghai Municipal Science and Technology Major Project under (Grant No.2019SHZDZX01) and the Shanghai talent program.

\bmhead{Data availability} All data underlying the results are available from the corresponding author upon reasonable request.

\bmhead{Competing interests} WPZ is an editorial board member for Quantum Frontiers and was not involved in the editorial review, or the decision to publish this article. All authors declare no competing interests.

\bibliography{sn-bibliography}

\end{document}